\documentclass[reprint,
 amsmath,amssymb,
 aps,
]{revtex4-2}
\usepackage{comment}

\usepackage{hyperref}
\usepackage{graphicx}
\usepackage{dcolumn}
\usepackage{bm}
\usepackage[utf8]{inputenc}
\usepackage[T1]{fontenc}
\usepackage{amsmath,amssymb,amsfonts}
\usepackage{physics}           
\usepackage{graphicx}          
\usepackage{bm}                
\usepackage{hyperref}          
\usepackage{xcolor} 
\usepackage{graphicx}
\usepackage{float}
\usepackage{microtype}         
\usepackage{mathtools}         
\usepackage{comment}  
\usepackage{subfigure}
\usepackage{booktabs} 

\begin{document}

\preprint{APS/123-QED}

\title{ Information-Theoretic Analysis of Weak Measurements and Their Reversal }

 \author{Luis D. Zambrano Palma}
  \email{ldzambra@tamu.edu}

\author{Yusef Maleki}
 \email{maleki@tamu.edu}

 \author{M. Suhail Zubairy}
 \email{zubairy@tamu.edu}

\affiliation{Institute for Quantum Science and Engineering, Texas A\&M University, College Station, Texas 77843, USA}

\date{\today}

\begin{abstract}

We study trade-off relations in information extraction from quantum systems subject to null-result weak measurements, where the absence of a detected photon continuously updates the system state. We present a detailed analysis of qubit and qutrit systems and investigate a general framework for a multilevel quantum system. We develop a dynamical characterization of null-result weak measurements that quantifies the information extracted over time, revealing the amount of the obtained information and also the rate of the information accumulation. The characterizations are obtained by examining the time-dependent evolution of the information theoretic quantities. More specifically, we consider Shannon entropy, mutual information, fidelity, and relative entropy to characterize the weak measurement dynamics. 
Our results provide an information theoretic analysis of the weak measurement process and highlight the dynamical nature of information extraction and reversibility in the weak measurement processes.

\end{abstract}


\maketitle

\section{Introduction}
There is an intrinsic connection between a quantum measurement and the information obtained from its possible outcomes. In quantum mechanics, a measurement projects the system onto an eigenstate associated with a definite eigenvalue, a process that lies at the foundation of the theory~\cite{nielsen2010quantum}. The selection of a single outcome from an initial superposition causes wavefunction collapse, thereby inducing irreversibility and destroying coherence. Such measurements correspond to projective or von Neumann, measurements~\cite{vonNeumann1955}. By contrast, in weak measurement the state does not fully collapse to an eigenstate of the system, allowing for partial information gain about the system~\cite{PhysRevB.60.5737}. In continuous-monitoring implementations, a record of (often null) outcomes gradually updates the state along quantum trajectories, allowing information to accumulate while nontrivial coherence is retained. Weak measurements therefore are important for measurement descriptions and also control of open quantum systems, including trajectory theory and measurement-based feedback~\cite{AlAmri_2011,Liao_2013}. Given the environment's deleterious role in eroding quantum coherence \cite{maleki2019stereographic,zurek2003decoherence} and entanglement \cite{maleki2018witnessing,haroche1998entanglement}, and in light of the need to control open quantum systems and to build noise-resilient protocols across quantum computing \cite{shor1995scheme,chuang1995quantum}, quantum communication \cite{bourennane2004decoherence,zhang2025quantum}, and quantum metrology \cite{escher2011general,maleki2018generating},  it is vital to explore the utility and implications of weak-measurement mechanisms.

The formulation of information-disturbance trade-off relations and correlation models for quantum measurements has become a central theme in both classical~\cite{PhysRevLett.90.050401,xi2013quantum,henderson2001classical,PhysRevLett.104.080501,wilde2015recoverability,PhysRevA.83.032114,PhysRevA.85.022124,PhysRevA.92.022114} and quantum~\cite{sun2024measurement,luo2010information,Lee2021,PhysRevA.93.052308,Luo2024,PhysRevA.53.2038,PhysRevLett.109.150402,PhysRevLett.100.210504} information theory. This line of research is driven by the need to understand how information is distributed during a measurement and how this distribution constrains the resulting disturbance, a connection that is frequently expressed using entropic measures. Experimental studies have confirmed this behavior, showing that entropy changes in generalized measurements follow trade-off relations consistent with classical information gain~\cite{Mancino2018}. An early step in this direction was taken by Ban, who studied photon counting as a form of continuous quantum measurement~\cite{BAN1997209}. Subsequent work established a mutual-information framework in which information is attributed both to the measurement outcomes and to the measurement-induced disturbance on the system~\cite{MasashiBan_1999,MasashiBan_19991,Ban1998}. More recently, weak measurements have been shown to admit probabilistic reversibility, where the state disturbance can be undone and the pre-measurement state recovered under suitable conditioning~\cite{PhysRevA.80.033838,PhysRevA.82.052323,PhysRevLett.82.2598,PhysRevLett.97.166805,Masashi11Ban_2001}. This line of work demonstrated that measurement back-action is not inherently irreversible, establishing weak measurement reversal as a tool for suppressing decoherence and enabling measurement-based quantum error correction~\cite{PhysRevLett.82.2598}. Important efforts have also focused on quantifying the information–disturbance balance, employing both quantum and classical forms of relative entropy~\cite{kullback1951information,kullback1997information}, which serve as fundamental measures of state distinguishability~\cite{RevModPhys.74.197,schumacher2000relative}. In this context, comparing the pre and post measurement states provides a quantitative means of characterizing how much the measurement has altered the system~\cite{RevModPhys.84.1655}, a question that becomes important when reversibility and recovery are likely. In this framework, Kuramochi and Ueda~\cite{Kuramochi2015} established a relation connecting the relative entropy change of a system observable to the relative entropy of the measurement outcomes, thereby linking state change to information gain within a single formalism.

In this work, we investigate how information, reversibility, and fidelity evolve over time in null-result weak measurements applied to finite-dimensional quantum systems. We formulate the problem within a classical information-theoretic framework, and track in detail the information gain in the null result, the mutual information, the relative entropy, the fidelity, and the reversal probability~\cite{PhysRevA.80.033838}. The latter two quantities are particularly significant, as they provide insight into the feasibility of reversing the effects of weak measurements and recovering the system’s coherence. We compute the time derivatives of the information gain, fidelity, and reversal probability, which provide instantaneous measures of information-extraction efficiency and coherence loss. These dynamical relations are obtained for qubits and qutrits and extend naturally to arbitrary \(N\)-level systems, where their characteristic features and limiting behaviors are identified. The results yield an ensemble-averaged, time-resolved characterization of information flow and irreversibility in null-result weak measurements.

This paper is organized as follows. In Sec.~\ref{sec2}, we introduce the framework of null-result during a weak measurements, define the key information-theoretic quantities, and investigate their dynamical evolution under continuous monitoring for two representative systems: a qubit and a qutrit. In Sec.~\ref{sec3}, we analyze the time derivatives of the central information–disturbance relations, characterizing the rates at which information is extracted and identifying the regimes where monitoring is most efficient. Finally, Sec.~\ref{sec4} summarizes our findings and outlines potential directions for future work.

\section{The scheme and informational analysis} \label{sec2}

\subsection{Multi-level setting}
We begin by considering the system in a superposition of photon-number (Fock) states inside the cavity. Equivalently, the dynamics may be regarded as those of a multi-level atom coupled to a single cavity mode. In the Fock basis, the state of the system is expressed as
\begin{equation}
    \ket{\psi} = \sum_{n=0}^{N} c_n \ket{n}.
\end{equation}
We suppose the cavity is being under continuous monitoring by an external detector. In a given time interval $t$, we assume that no detection event (i.e., no photon decay) is recorded, corresponding to a null-result outcome. The associated measurement operator is given by ~\cite{ueda1989probability}  
\begin{equation}
    M_{0} = e^{-\gamma t \hat{n}}, 
    \label{wmo}
\end{equation}
where $\hat{n}$ is the photon-number operator and $\gamma$ denotes the effective decay rate. Therefore, after a weak measurement with a null result, the field state evolves to
\begin{equation}
\ket{\psi'} = \frac{M_0 \ket{\psi}}{\sqrt{\bra{\psi} M_0^{\dagger} M_0 \ket{\psi}}}. \label{mse}
\end{equation}
In the present case, the system state conditioned on a null detection within a time interval $t$ takes the form
\begin{equation}
    \ket{\psi'} = \frac{\sum_{n=0}^{N} c_n e^{-n \gamma t} \ket{n}}{\sum_{m=0}^{N} |c_m|^2 e^{-2 m \gamma t}},
\end{equation}
where $\gamma$ is the decay rate, $t$ is the evolution time. The probability of obtaining a null result at time $t$ is
\begin{equation}
    p(y_{0}) = \bra{\psi} M_0^{\dagger} M_0 \ket{\psi} 
    = \sum_{n=0}^{N} |c_n|^2 e^{-2 n \gamma t}. \label{nre}
\end{equation}
Consequently, the conditional probability of the photon number outcome being $n$, given a null detection, is
\begin{equation}
    p(x_n|y_{0}) = 
    \frac{p(y_{0}|x_n)\, p(x_n)}{p(y_{0})}
    = \frac{|c_n|^2 e^{-2 n \gamma t}}
    {\sum_{m=0}^{N} |c_m|^2 e^{-2 m \gamma t}},
    \label{eq:nullpost1}
\end{equation}
where $p(x_n) = |c_n|^2$ is the prior distribution and 
\( p(y_{0}|x_n) = \bra{n} M_0^{\dagger} M_0 \ket{n} 
    = e^{-2 n \gamma t}.\)

\subsection{Shannon Entropy and Mutual Information}
Once the null-result dynamics are defined, we now turn to a quantitative characterization of how information evolves under weak measurement. To this end, we adopt an information-theoretic framework. The collapse of a quantum state into a photon-number eigenstate $\ket{n}$ is represented by a classical probability distribution $p(x_n)$ associated with the random variable $X$, while the absence of detector clicks is described by a distribution $p(y_0)$ corresponding to the random variable $Y$. Within this framework, we compare the Shannon entropy of the initial distribution with that of the posterior distribution conditioned on a null result. The initial Shannon entropy is defined as~\cite{shannon1948claude}
\begin{equation}
    H(X) = -\sum_{n=0}^{N} p(x_n)\,\log_{2} p(x_n).
\end{equation}
Following a null result, the conditional Shannon entropy is defined as
\begin{equation}
    H(X|y_0) = -\sum_{n=0}^{N} p(x_n|y_0)\,\log_{2} p(x_n|y_0).
\end{equation}
The information gain associated with a null outcome can be expressed as
\begin{equation}
\begin{aligned}
I(0) &= H(X) - H(X|y_0) \\
     &= \sum_{n=0}^{N} \big[\, p(x_n)\,\log_{2} p(x_n)
        - p(x_n|y_0)\,\log_{2} p(x_n|y_0) \,\big].
\end{aligned}
\end{equation}
which quantifies the change in uncertainty when conditioning on the absence of decay.  
The mutual information $I(X:Y)$ is obtained by averaging the information gain over all possible outcomes  
\begin{equation}
    I(X:Y) = \sum_{k} p(y_k)\, I(k),
\end{equation}
where $p(y_k)$ is the probability of detector outcome $y_k$ (with the null result probability $p(y_0)$ defined in Eq.~\eqref{nre}).  
This mutual information represents the expected reduction in uncertainty about the system induced by the measurement process, and thus characterizes the net information extracted by the detector.

\subsection{Reversal Probability}

Normally, a projective measurement reveals the full information of an unknown system through the instantaneous collapse of the wave function into a specific eigenstate. In contrast, a null result in a weak measurement gradually extracts information about the system while only partially disturbing its state. Considerable effort has been devoted to investigating whether such measurement-induced decoherence can be undone. In particular, we follow the approach of Sun \emph{et al.}~\cite{PhysRevA.80.033838}, who analyzed the reversal success probability for the weak measurement protocol involving two successive weak measurements. In this framework, the success probability is defined as
\begin{equation}
    P_{\mathrm{rev}}(t) = \frac{e^{-2\gamma N t}}{p(y_0)},
\end{equation}
where \(N\) is the maximum excitation number and \(p(y_0)\) is the probability of no detection up to time \(t\). This quantity gives the conditional probability of restoring the system to its initial state and highlights the trade-off between information gain and coherence recovery, an important theme in quantum error correction and state-protection strategies~\cite{gottesman2009introduction}.

\subsection{Fidelity}

In addition to the mutual information, we quantify the similarity between the initial distribution \(p(x_n)\) and the updated null-result distribution \(p(x_n|y_0)\) using the classical fidelity~\cite{nielsen2010quantum},
\begin{equation}
       F(t) = \sum_{n=0}^{N} \sqrt{p(x_n)\,p(x_n|y_0)} 
       = \frac{1}{\sqrt{p(y_0)}} \sum_{n=0}^{N} p(x_n)\,e^{-\gamma n t}.
\end{equation}
The fidelity provides a time-dependent measure of the disturbance induced by the null-result measurement, expressed as the overlap between the original and the updated probability distributions.

\subsection{Relative Entropy}

So far, we have discussed the mutual information and fidelity, both of which serve as valuable indicators of information loss and coherence degradation under weak measurements. However, these quantities alone do not fully capture the asymmetry between the null-result evolution of the system and the dynamics required to reverse it. To address this, we introduce the concept of relative entropy~\cite{kullback1951information,kullback1997information}, which provides a quantitative measure of the distinguishability, or asymmetry, between two probability distributions, defined as ~\cite{kullback1951information,kullback1997information}
\begin{equation}
    D (p_n \| q_n ) = \sum_{n=0}^{N} p_n \log_2 \frac{p_n}{q_n}.
\end{equation}
In the context of weak measurements with a null-result outcome, we use this measure to compare the initial distribution \(p(x_n)\) with the post-selected (null-conditioned) distribution \(p(x_n|y_0)\). In particular, the relative entropy from \(p(x_n|y_0)\) to \(p(x_n)\) defines a different quantity
~\cite{kullback1951information,kullback1997information}
\begin{equation}
    D(p(x_n|y_0)\,\|\,p(x_n)) 
    = \sum_{n=0}^{N} p(x_n|y_0)\,\log_2 \frac{p(x_n|y_0)}{p(x_n)}.
\end{equation}

\subsection{Information Quantifiers and Weak Measurement}

\begin{figure}[H]
  \centering
  \includegraphics[width=1\linewidth]{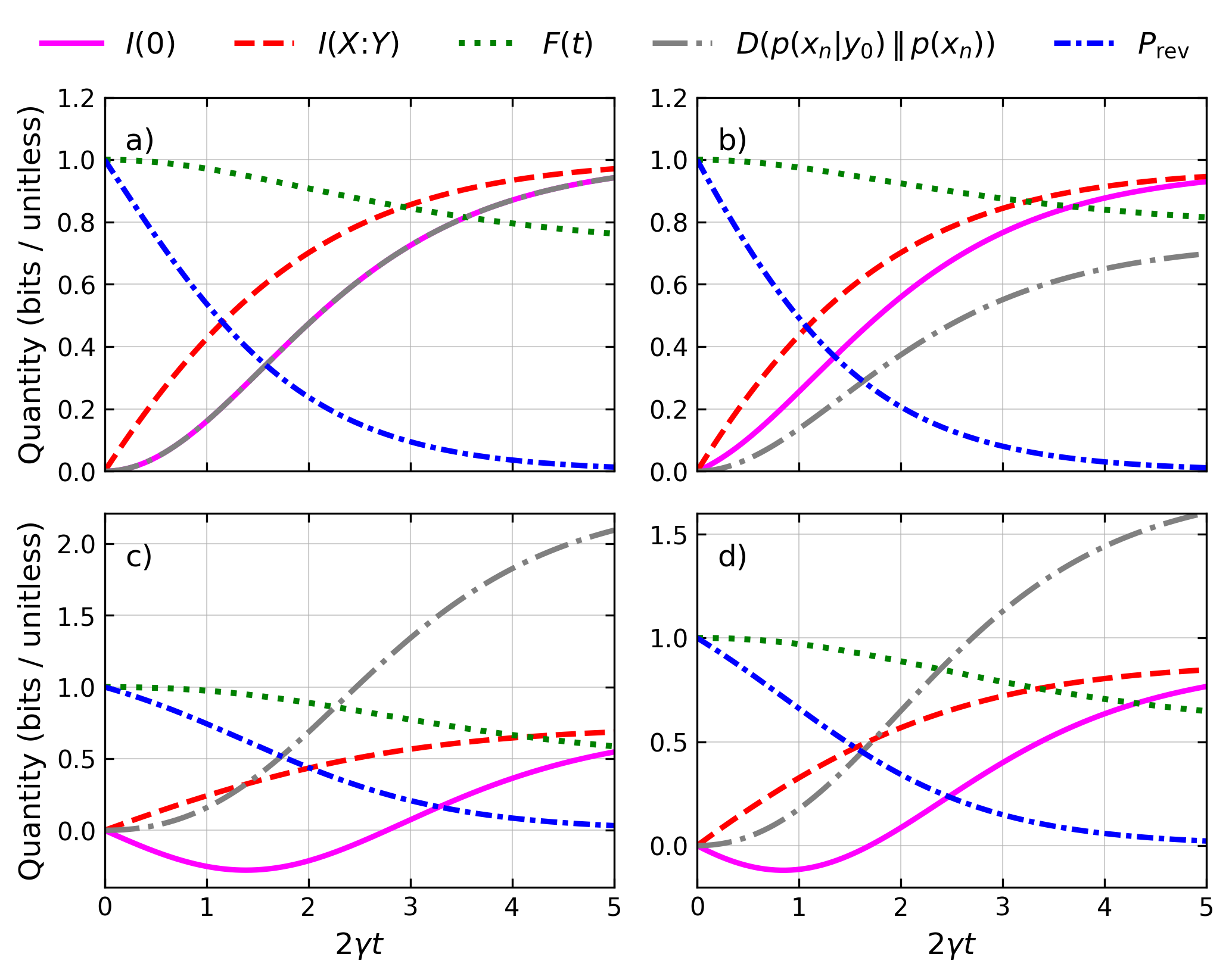}
\caption{
    Information-theoretic quantities as functions of $2\gamma t$ for four different prior distributions for the state \(\ket{\psi}=c_0\ket{0}+c_1\ket{1}\). 
    Each panels correspond to different priors: a) $p(x_n) = [1/2,\,1/2]$, b) $p(x_n) = [0.6,\,0.4]$, c) $p(x_n) = [0.2,\,0.8]$, and d) $p(x_n) = [0.3,\,0.7]$.
  }
  \label{fig1}
\end{figure}

\begin{figure}[H]
  \centering
  \includegraphics[width=1\linewidth]{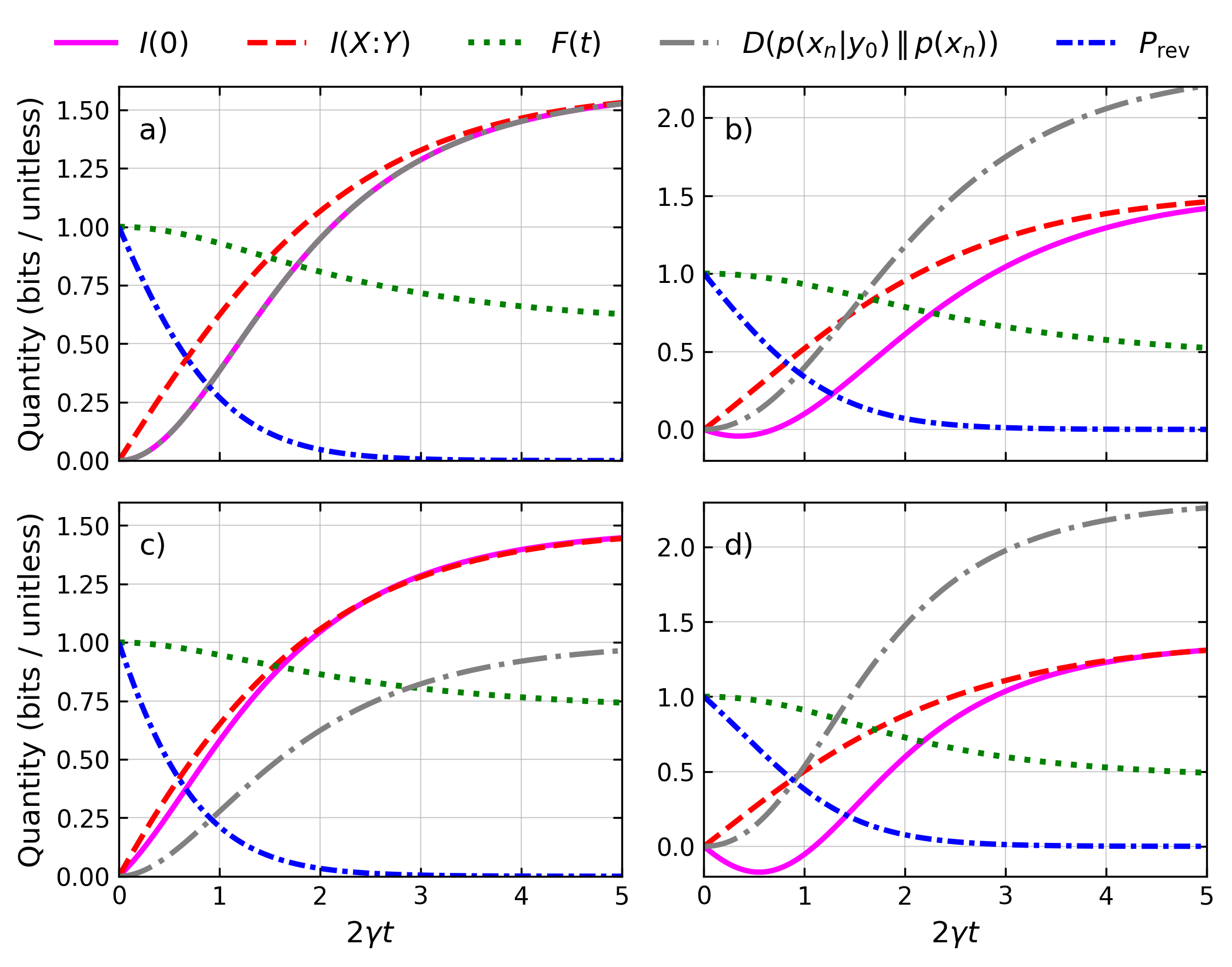}
\caption{
    Information-theoretic quantities as functions of $2\gamma t$ for four different prior distributions for the state \(\ket{\psi}=c_0\ket{0}+c_1\ket{1}+c_2\ket{2}\). 
    Each panel corresponds to different priors: a) $p(x_n) = [1/3,\,1/3,\,1/3]$, b) $p(x_n) = [0.2,\,0.4,\,0.4]$, c) $p(x_n) = [0.5,\,0.3,\,0.2]$, and d) $p(x_n) = [0.2,\,0.2,\,0.6]$.
  }
  \label{fig2}
\end{figure}

In the previous subsections, we introduced relevant measures that quantify the information gained from a null-result weak measurement. Figs.~\ref{fig1} and~\ref{fig2} illustrate the defined information quantities as functions of the scaled time \(2\gamma t\): the information gain, the mutual information, the fidelity, the reversal probability, and the relative entropy, for a qubit and qutrit under null-result monitoring with four different initial probability distributions. This comparison demonstrates how the measures evolve and highlights the distinct insights each provides into the trade-offs between information gain, coherence, and irreversibility.

We present the analysis of the uniform distribution in Fig.~\ref{fig1}(a). Here, the information gain \(I(0)\) increases monotonically, reflecting the continuous accumulation of information due to a constant null-result updated. As it is expected in weak measurement, the detector gradually reduces uncertainty by eliminating contributions from decaying components. The relative entropy exhibits the same monotonic increase, and in this case, both quantities converge to the same value. This coincidence indicates that, at early times, the post-selected distribution remains close to the prior, making the system more easily reversible. This interpretation is consistent with the fidelity, which stays close to unity over the same time range \(t \in (0,1)\), confirming that the null-result state retains a high overlap with the initial state. The mutual information also increases monotonically and approaches to a saturation value that nearly coincides with the saturation value of the other information measures. This convergence suggests that, in the long-time limit, the null-result state becomes perfectly distinguishable from the initial state, and nearly all accessible information has been extracted.

In addition, in the case of the nonuniform priors distribution shown in Fig.~\ref{fig1}(b)–(d), the four measures separate more clearly and reveal how the prior bias controls the measures dynamics. In panel (b), the relative entropy \(D(p(x_n|y_0)\|p(x_n))\) grows the fastest and reaches a larger asymptotic value than the other quantities, indicating a larger reduction in distribution by the null-result update. The information gain \(I(0)\) increases more slowly than the mutual information \(I(X{:}Y)\) at early times (since \(I(X{:}Y)\) averages over all possible outcomes), but both eventually approach the same limit set by the ground-state prior \(p_0\) [i.e., \(\sim H(x_n)\)]. The fidelity \(F(t)\) decays from unity to a lower value than in the uniform case, reflecting the stronger dependence on the initial probability distribution.

We continue with Fig.~\ref{fig1}(c), which corresponds to a prior closer to the uniform case. Here, \( I(0)\) and \(I(X{:}Y)\) remain close to each other throughout the entire evolution and saturate at nearly the same value. In this case, the information gain and mutual information demonstrate similar behavior. In contrast, the relative entropy increases more slowly in this case. This behavior indicates that the null-result update does not significantly modify the distribution, keeping the posterior relatively close to the prior for a longer interval.

In Fig.~\ref{fig1}(d), the influence of the prior distribution is most pronounced. The fidelity drops rapidly and saturates to the same value as in Fig.~\ref{fig1}(b), while the relative entropy becomes the dominant curve in both slope and final magnitude, indicating that the null-result posterior is highly asymmetric with respect to the prior. By contrast, \(I(X{:}Y)\) remains strictly positive and stays above \(I(0)\) during the transient before the two measures converge. Taken together, the behavior from Fig.~\ref{fig1}(b)-(d) shows that a faster decay of \(F(t)\) and a larger final value of \(D\) signal progressively stronger redistribution of probability weight toward lower photon numbers under the null-result update, while the long-time limits of both \(I(0)\) and \(I(X{:}Y)\) remain fixed by the same prior-dependent bound.

 An important feature appears in the Figs.~\ref{fig1}(b) and (d) where in small time windows the information gain \(I(0)\) becomes negative. As discussed in Ref.~\cite{d2003heisenberg}, negative values can arise because \(I(0)\) is an unaveraged quantity: for certain priors and outcomes, the posterior distribution may carry more uncertainty than the prior. In such cases, the measurement effectively increases uncertainty rather than reducing it.

In Fig.~\ref{fig2},  we present the information–disturbance quantities for the qutrit case, analyzed under four different prior distributions. The overall behavior is similar to that of the qubit: for the uniform prior, the information gain coincides with the relative entropy, whereas for nonuniform priors, the two measures show a significant difference. A notable distinction is that both fidelity and reversal probability drop more rapidly than the qubit case, reflecting the enhanced changes in higher dimensions. The presence of the additional level accelerates coherence loss and reduces the chance of reversibility, which can be interpreted as the detector having a higher likelihood of registering a leakage photon when more excitations are available. The mutual information presents a similar qualitative behavior to the qubit case and remains strictly non-negative for all priors. In all scenarios, the three quantities exhibit saturation, with their asymptotic values strongly influenced by the prior distribution, which determines how much information can be extracted through null-result weak measurements. In Figs.~\ref{fig2}(b) and (d) the information gain \(I(0)\) even dips to negative values at short times, reflecting situations in which the posterior carries more uncertainty than the prior, as it was discussed before.

\begin{table}[ht]
\caption{Threshold times ($2\gamma t$)* for fidelity, reversal probability, and information gain in the qubit case.}
{\begin{tabular}{@{}cccc@{}}
\toprule
Panel & $F < 90\%$ & $P_{\rm rev} < 50\%$ & $I(0) > 90\% I_{\max}$ \\ \colrule
(a)    & 2.124 & 1.104 & 3.813 \\
(b)   & 2.475 & 0.987 & 3.562 \\
(c)    & 1.923 & 1.806 & 4.649 \\
(d)    & 1.873 & 1.472 & 4.348 \\
\botrule
\end{tabular}}
\label{tab1}
\end{table}

\begin{table}[ht]
\caption{Threshold  times ($2\gamma t$)* for fidelity, reversal probability, and information gain in the qutrit case.}
{\begin{tabular}{@{}cccc@{}}
\toprule
Panel & $F<90\%$ & $P_{\rm rev}<50\%$ & $I(0)>90\% I_{\max}$ \\
\\ \colrule
(a)    & $ 1.254$ & $ 0.585$ & $ 3.445$ \\
(b)    & $ 1.237$ & $ 0.702$ & $ 3.913$ \\
(c)    & $ 1.555$ & $ 0.485$ & $ 3.127$ \\
(d)   & $ 1.070$ & $0.786$ & $ 3.662$ \\
\botrule
\end{tabular}}
\label{tab2}
\end{table}

Next, we identify characteristic times that distinguish the relevant informational and dynamical regimes in Tables~\ref{tab1} and \ref{tab2}. The first threshold corresponds to fidelity dropping below $90\%$, which for the qubit occurs at $(2\gamma t)^* \approx 1.87$–$2.12$, but for the qutrit occurs already at $(2\gamma t)^* \approx 1.0$–$1.25$, demonstrating faster coherence loss in higher dimensions. In Table~\ref{tab1}, we show that, for the qubit, the reversal probability $P_{\mathrm{rev}}$ falls below $50\%$ within $(2\gamma t)^* \approx 0.98$–$1.80$, indicating that reversibility is rapidly compromised. By contrast, In Table~\ref{tab2} we show that the qutrit reaches the same threshold even earlier, at $(2\gamma t)^* \approx 0.58$–$0.78$, leaving less opportunity for recovery. Finally, the information gain reaches $90\%$ of its maximum at $(2\gamma t)^* \approx 3.5$–$4.6$ for the qubit and $(2\gamma t)^* \approx 3.1$–$3.9$ for the qutrit, indicating slightly faster information extraction in the higher-dimension cases Overall, these results highlight that increasing system dimension accelerates both the onset of coherence loss and the reduction of reversibility, thereby diminishing the prospects for recovery under null-result weak measurements.

\section{Instantaneous rates of information gain and irreversibility}  \label{sec3}

To complement the analysis of informational quantities, we now examine the
temporal rates at which information is acquired and reversibility is lost during null-result weak measurements. This provides a time-resolved characterization of the measurement-induced dynamics, extending the discussion of trade-offs between information gain, coherence, and reversibility at fixed times.

The instantaneous rate of information gain associated with the null outcome is defined with respect to the scaled time $\tau = 2\gamma t$ as
\begin{equation}
    \dot{I}(0,\tau) = \frac{d}{d \tau} I(0,\tau).
\end{equation}
Using the conditional distribution $p(x_n|y_0)$, this can be written as
\begin{equation} \dot{I}(0,\tau) = -\sum_{n=0}^{N} p(x_n|y_0)\, \left(n -\frac{\sum_{m} m\, p(x_m)\, e^{-m\tau}}{p(y_0)}\right) \log_2 p(x_n|y_0). 
\end{equation}
Therefore, we find
\begin{equation}
    \lim_{\tau\rightarrow 0} \dot{I}(0) =-\sum_{n=0}^{N} p(x_n)\,
    \left(n -\sum_{m=0}^{N} m\, p(x_m)\right)\log_2 p(x_n), \label{eq17}
\end{equation}
which reduces to
\begin{equation}
    \lim_{\tau\rightarrow 0} \dot{I}(0) =\begin{cases}
-\,p(x_0)\, p(x_1)\,\log_2 \dfrac{p(x_1)}{p(x_0)}, & \text{qubit}, \\
\\
\langle n\rangle\,p_0\log_2 p_0
-(1-\langle n\rangle)\,p_1\log_2 p_1
\\-(2-\langle n\rangle)\,p_2\log_2 p_2, & \text{qutrit},
\end{cases}
\end{equation}
with $\langle n \rangle=p(x_1)+2\,p(x_2)$, and
\begin{equation}
    \lim_{\tau\rightarrow \infty} \dot{I}(0) = 0.
\end{equation}

Since information gain is accompanied by reduced reversibility, we also quantify the rates at which fidelity and reversal probability change. The instantaneous rate of fidelity is defined by
\begin{equation}
    \dot{F}(\tau) = \frac{d}{d\tau} F(\tau),
\end{equation}
with explicit form
\begin{equation}
    \dot{F}(\tau) = \frac{1}{2}\left[
    \frac{F(\tau)\,\sum_{n=0}^{N} n\, p(x_n)\, e^{- n \tau}}{p(y_0)}
    -\sum_{n=0}^{N} n\,\sqrt{p(x_n)\,p(x_n|y_0)}\right],
\end{equation}
satisfying
\begin{equation}
    \lim_{\tau\rightarrow 0} \dot{F} =
    \lim_{\tau\rightarrow \infty} \dot{F} = 0. \label{eq22}
\end{equation}
Similarly, the rate of change of the reversal probability is
\begin{equation}
    \dot{P}_{\text{rev}}(\tau) = \frac{d}{d \tau} P_{\text{rev}}(\tau),
\end{equation}
with
\begin{equation}
    \dot{P}_{\text{rev}}(\tau) =  P_{\text{rev}}(\tau)\left[
    \frac{\sum_{n=0}^{N} n\, p(x_n) \, e^{- n \tau}}{p(y_0)}-N \right].
\end{equation}
The limiting behaviors are
\begin{equation}
    \lim_{\tau\rightarrow0} \dot{P}_{\text{rev}} =
    \langle n \rangle - N
    =\begin{cases}
p(x_1)-1, & \text{qubit}, \\[6pt]
p(x_1)+2p(x_2)-2, & \text{qutrit},
\end{cases}
\end{equation}
where $\langle n \rangle=\sum_{n=0}^{N} n\, p(x_n)$, and
\begin{equation}
    \lim_{\tau\rightarrow \infty} \dot{P}_{\text{rev}} = 0.
\end{equation}

\begin{figure}[H]
  \centering
  \includegraphics[width=1\linewidth]{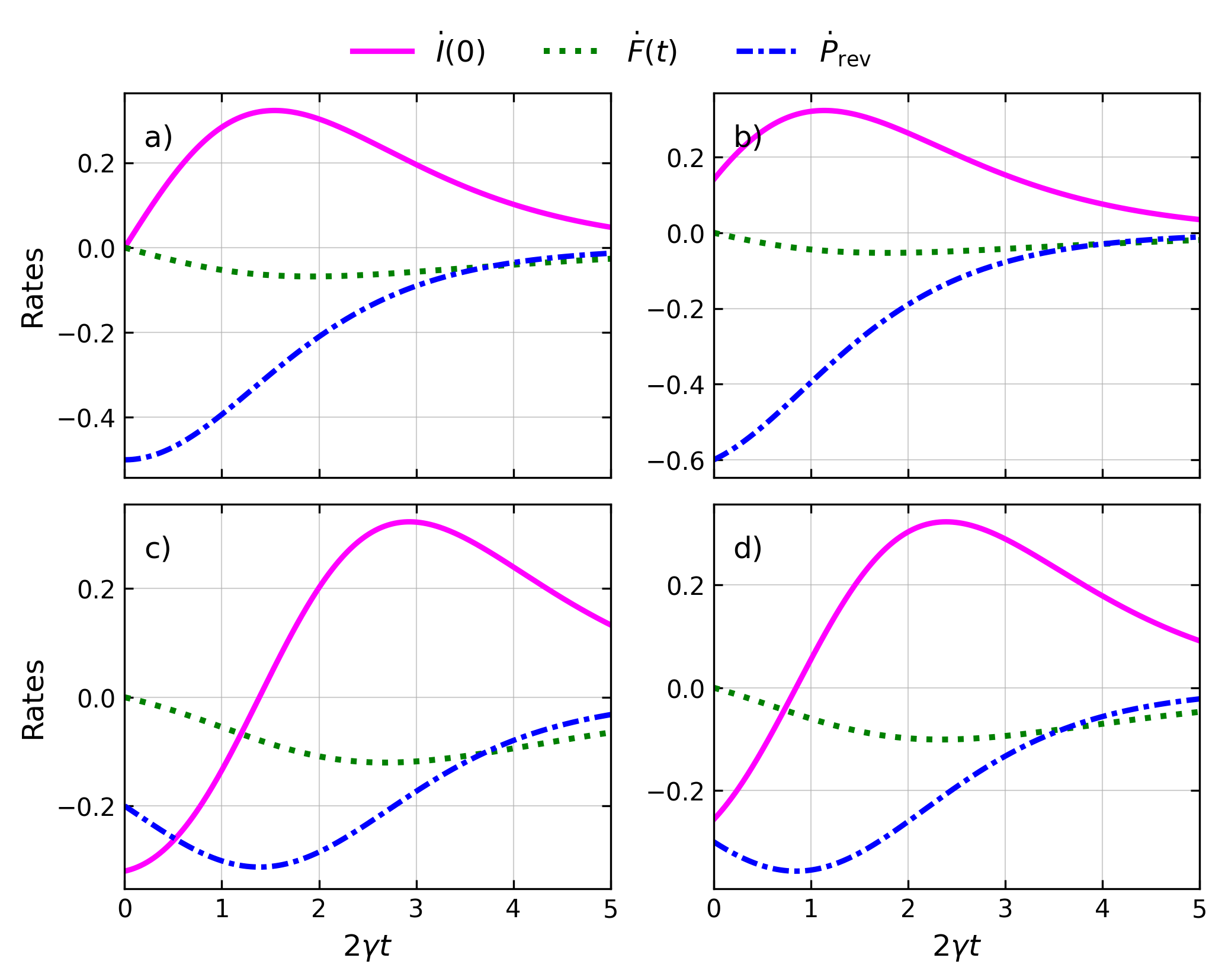}
\caption{
    Instantaneous rates of change of information gain, fidelity, and reversal probability for a qubit under null-result monitoring as functions of $2\gamma t$. Each panel corresponds to a different prior distribution: a) $p(x_n) = [1/2,\,1/2]$, b) $p(x_n) = [0.6,\,0.4]$, c) $p(x_n) = [0.2,\,0.8]$, and d) $p(x_n) = [0.3,\,0.7]$.
  }
  \label{fig3}
\end{figure}

\begin{figure}[H]
  \centering
  \includegraphics[width=1\linewidth]{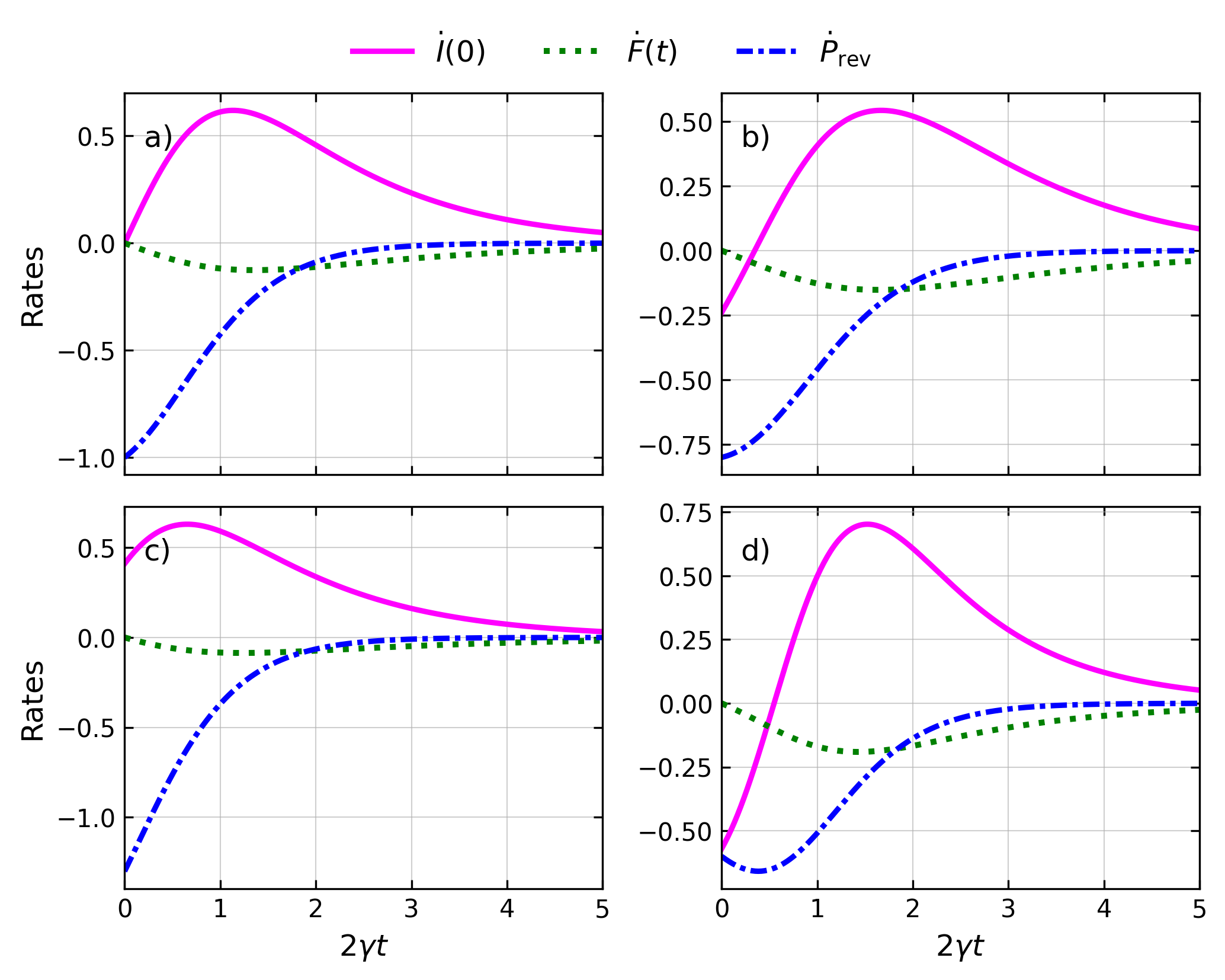}
\caption{
    Instantaneous rates of change of information gain, fidelity, and reversal probability for a qutrit under null-result monitoring as functions of $2\gamma t$. Each panel corresponds to a different prior distribution: a) $p(x_n) = [1/3,\,1/3,\,1/3]$, b) $p(x_n) = [0.2,\,0.4,\,0.4]$, c) $p(x_n) = [0.5,\,0.3,\,0.2]$, and d) $p(x_n) = [0.2,\,0.2,\,0.6]$.
  }
  \label{fig4}
\end{figure}

Now, to gain insights into the rates at which information is acquired and reversibility is lost, we present in Figs.~\ref{fig3} and~\ref{fig4}  the instantaneous rates $\dot{I}(0,\tau)$, $\dot{F}(\tau)$, and $\dot{P}_{\mathrm{rev}}(\tau)$ for qubits and qutrits, respectively, under the same four prior distributions considered in the section above.

In Fig.~\ref{fig3}, the rate $\dot{I}(0,\tau)$ quantifies the instantaneous speed of information acquisition. As shown in Eq.~\ref{eq17}, its initial value is determined by the prior probability distribution. In particular, for the biased priors of Fig.~\ref{fig3}(c)–(d) this rate is negative, indicating that information extraction decreases from the outset. Although seemingly counterintuitive, this behavior reflects an initial increase in uncertainty: instead of revealing information, the chosen prior results in a temporary reduction of information gain. By contrast, for the uniform distribution in Fig.~\ref{fig3}(a) the initial rate vanishes, while in Fig.~\ref{fig3}(b) it is positive. In all cases, the rate subsequently increases, reaching a peak that marks the maximum velocity of information extraction, and then decreases again. This behavior is consistent with the characteristic times at which $90\%$ of the information is acquired, as reported in Tables~\ref{tab1} and ~\ref{tab2}. At long times, the rate tends to zero, reflecting the saturation of the information gain shown in Figs.~\ref{fig1} and \ref{fig2}.

In contrast, the rate of fidelity $\dot{F}(\tau)$ remains negative for all times, reflecting the continuous loss of similarity between the initial state and the null-result state. As described in Eq.~\ref{eq22}, the rate always starts from zero. Its minimum corresponds to the point of the most rapid fidelity decay, which coincides with the regime of maximum information extraction, reinforcing the intrinsic link between information gain and disturbance. At long time, the rate of fidelity approaches zero, consistent with the eventual saturation of the fidelity.

In addition, the rate $\dot{P}_{\mathrm{rev}}(\tau)$ starts at different values depending on the prior distribution, indicating that the system begins to lose recoverability immediately. This rate reaches its maximum magnitude at the very beginning, as seen in Table~\ref{tab1}, and remains negative throughout. Its overall behavior is similar to that of $\dot{F}(\tau)$, since both quantities decay monotonically, reflecting a continuous loss of coherence and reversibility. At long times, the rate approaches zero, consistent with the saturation of the reversal probability. This behavior underscores the irreversibility introduced by continuous conditioning on the absence of decay.  

The trends observed for the qubit case in Fig.~\ref{fig3} are also present for the qutrit in Fig.~\ref{fig4}, where the increase in the number of level slightly enhances both the rate of information extraction and the rate of recoverability loss. In particular, the initial drop of $\dot{P}_{\mathrm{rev}}(\tau)$ is steeper for the qutrit, indicating that higher-dimensional systems experience faster irreversibility under null-result monitoring. Moreover, the regime of rapid information acquisition occurs earlier than in the qubit case, consistent with the results summarized in Table~\ref{tab1}.

\section{Conclusion} \label{sec4}

In summary, this work quantitatively characterizes how the structure of the initial quantum state constrains information acquisition in continuous null-result weak measurements. We use a systematic comparison of qubit and qutrit cases, with an extension to a general multilevel framework, characterizing the relation between state reduction and the reversibility in various settings.
Tracking the evolution of Shannon entropy, mutual information, fidelity, and relative entropy allowed us to quantify the amount of the obtained information, and also the rate of the information gain.
 This dynamical characterization highlights the connection of information extraction and reversibility. 
Our results reinforce a central characteristics of quantum measurement theory and control where the reversibility of information is not solely determined by the measurement process but is also linked to the statistical structure of the initial state. Also, the framework developed here provides a framework to compare these trade-offs across different system dimensions and measurement strengths.

\bibliography{refs}

\end{document}